\RequirePackage{fix-cm}
\documentclass{svjour3}                     
\smartqed  
\usepackage{amssymb}
\usepackage{graphicx}
\usepackage{pdflscape}
\usepackage{multirow} 
\usepackage{rotating}
\usepackage{url} 
\usepackage{longtable}
\usepackage{lscape} 

\usepackage{fancyvrb}
\usepackage{fvextra}
\usepackage{float}
\usepackage{array}
\usepackage{tabularx}
\usepackage{multirow}
\usepackage{xcolor}
\usepackage{verbatim} 
\usepackage{alltt}
\usepackage{listings}
\usepackage{caption} 
\usepackage{colortbl}
\usepackage{lscape}

\usepackage[colorlinks=true,linkcolor=blue,urlcolor=black,bookmarksopen=true]{hyperref}
\hypersetup{breaklinks=true}
\usepackage[open,openlevel=3]{bookmark}
\usepackage{tablefootnote}
\usepackage{comment}

\newboolean{showcomments}
\setboolean{showcomments}{true} 
\ifthenelse{\boolean{showcomments}}
  {\newcommand{\nb}[2]{
    \fcolorbox{gray}{yellow}{\bfseries\sffamily\scriptsize#1}
    {\sf\small$\blacktriangleright$\textit{#2}$\blacktriangleleft$}
  }
  
  }
  {\newcommand{\nb}[2]{}
  
  }

\usepackage{hyperref}
\usepackage{makecell}

\usepackage{lineno}

\usepackage{booktabs}

\begin{document}

\title{Harmonizing DevOps Taxonomies
}
\subtitle{Theory Operationalization and Testing}

\titlerunning{Harmonizing DevOps Taxonomies -- Theory Operationalization and Testing}        

\author{Isaque Alves \and  Jorge P\'erez \and
        Jessica D\'iaz \and   Daniel L\'opez-Fern\'andez  \and Manuel Pais     \and Fabio Kon \and Carla Rocha         
}


\institute{
              I. Alves \and F. Kon. 
              Department of Computer Science, University of S\~ao Paulo.  S\~ao Paulo, Brazil.
              \email{\{isaque.alves, kon\}@ime.usp.br}
            \\
              J. D\'iaz \and D. L\'opez\and J. P\'erez. 
              Universidad Polit\'ecnica de Madrid. Departamento de Sistemas Inform\'aticos. ETSI Sistemas Inform\'aticos, Madrid, Spain.
              \email{\{yesica.diaz, daniel.lopez@upm.es,  jorgeenrique.perez\}@upm.es} 
             \\
              M. Pais. 
              Independent \& FlowOnRails SL, Madrid, Spain.
              \email{manuel@teamtopologies.com}
            \\
              C. Rocha.
              Universidade de Brasília. Faculdade do Gama. Gama, Brazil.
              \email{caguiar@unb.br}           \\
          }

\date{Received: date / Accepted: date}

\maketitle
\begin{abstract}
\parindent=0mm \color{white}.\color{black}

\textbf{Context:} 
DevOps responds the growing need of companies to streamline the software development process and, thus, has experienced widespread adoption in the past years. However, the successful adoption of DevOps requires companies to address significant cultural and organizational changes. Understanding the organizational structure and characteristics of teams adopting DevOps is key, and comprehending the existing theories and representations of team taxonomies is critical to guide companies in a more systematic and structured DevOps adoption process. As there was no unified theory to explain the different topologies of DevOps teams, in previous work, we built a theory to represent the organizational structure and characteristics of teams adopting DevOps, harmonizing the existing knowledge.
\textbf{Objective:}  In this paper, we expand the theory-building in the context of DevOps Team Taxonomies. Our main contributions are presenting and executing the Operationalization and Testing phases for a continuously evolving theory on DevOps team structures. 
\textbf{Method:} We operationalize the constructs and propositions that make up our theory to generate empirically testable hypotheses to confirm or disconfirm the theory. Specifically, we focus on the research operation side of the theory-research cycle: identifying propositions, deriving empirical indicators from constructs, establishing testable hypotheses, and testing them.
\textbf{Results:} We performed the operationalization and testing of the DevOps Team Taxonomies Theory, which resulted in an empirically verified and trustworthy theory.  Our theory has 28 propositions representing this model that map properties to the constructs of our theory. The operationalization generated 34 testable hypotheses, and we thoroughly tested 11 of them.
\textbf{Conclusions:} The testing has proved the effectiveness of the theoretical framework, while the operationalization of the constructs has enhanced the initial theoretical framework.

\keywords{DevOps Team Structures \and DevOps Taxonomies \and Theory Operationalization \and Theory Testing}
\end{abstract}

\section{Introduction}
\label{SecIntroduction}



Investment in the implementation and adoption of DevOps   has increased in the last few years, and surveys show DevOps is on the rise. However, successful DevOps adoption requires a significant cultural and organizational change in the IT departments of companies. Understanding the organizational structure and characteristics of teams adopting DevOps is key to establishing a successful adoption strategy and metrics. Comprehending existing theories and representations of team topologies is essential for guiding companies in a more systematic and structured DevOps adoption process.

Previous works have attempted to describe one or several team structures and how their characteristics impact the software delivery performance, \cite{leite2021,lopez2021,luz2019,macarthy2020,nybom2016,stateofdevops,shahin2017,zhou2022cross}. However, they do not adopt a shared vocabulary or concepts to describe each team structure. 
We proposed in  \cite{diaz2022} to harmonize the DevOps taxonomies from the existing studies to represent the organizational structure and characteristics of teams adopting DevOps. We collected and analyzed existing studies on DevOps teams taxonomies and conducted a Grounded Theory (GT) study \cite{charmaz:2014,diaz2021GT,glaser:1967}. The result was a theory that harmonizes DevOps Team Taxonomies in software-producing organizations\footnote{Hereinafter, we will refer to the DevOps Team Taxonomies Theory as T3.}; i.e., we proposed a model that harmonizes existing team taxonomies in the way organizations structure their development and infrastructure/operation teams around the DevOps culture and practices~\cite{diaz2022}. 



 According to the reference theory-building frameworks \cite{dubin:1978},   \cite{lynham2002}, and \cite{sjoberg:2008}, once we propose a theory, it is necessary to operationalize it as well as to subject it to rigorous testing to provide evidence for or against the theory. Operationalization and testing are the objectives of the present study. We present a methodological framework for continuous theory-building based on the available frameworks,  the ones by Dubin \cite{dubin:1978}, and Lynham \cite{lynham2002}, which were subsequently synthesized in the software engineering building-theory framework by Sj{\o}berg et al. \cite{sjoberg:2008}. We propose procedures for Theory Operationalization and Testing, and we conduct a confirmatory case study to provide empirical evidence that confirms/refutes our theory.
Thus, the contribution described in this paper is part of broader research investigating DevOps team topologies.  Extensive research data is made available as open data\footnote{\scriptsize{\url{https://github.com/alvesisaque/devops_taxonomies.github.io}}} so that other researchers can verify and extend this work in new directions.

This paper is structured as follows. Section~\ref{sec:background} presents an overview of the work related to the characterization of DevOps team structures and an overview of our DevOps Team Taxonomies Theory (T3). Section~\ref{sec:methodology} describes and justifies the methodologies employed in this research. In Section~\ref{sec:theory-operationalization} we address the operationalization of both constructs and propositions that make up T3 \cite{diaz2022}. Section~\ref{sec:usecase} describes the confirmatory case study: data collection, data processing, and data analysis. Section~\ref{sec:hypoheses-validation} validates the hypotheses developed in Section~\ref{sec:theory-operationalization} against the information generated by the case study analysis and describes the confirmatory and refutational aspects of the theory. In section~\ref{sec:discussion}, we discuss our major results. The threats to the validity of this research and the strategies implemented to mitigate them are described in Section~\ref{sec:threats}. Finally, Section~\ref{sec:conclusion} presents our main conclusions of this research and further work.

\section{Background: DevOps Team Taxonomies}
\label{sec:background}

This section discusses some of the key concepts around DevOps team taxonomies, which served as a foundation for T3, as well as the theory-based case study that will serve for testing our theory.

\subsection{DevOps Team Taxonomies}
\label{sec:back1}

Over the last decades, organizations have become more interested in the DevOps benefits \cite{diaz2021many} and have looked for ways to adopt and handle all the concepts and techniques around this concept. 
As a consequence of DevOps adoption, companies change their team structures and relationships to address organizational silos and conflicts, ownership sharing, and cultural values, among others. 

One of the first papers analyzing team structures is the work by Shahin et al. \cite{shahin2017}. They conducted a mixed-method empirical study that collected data from 21 interviews in 19 organizations and a survey with 93 practitioners. They identified four common types of team structures: \textbf{separate Dev and Ops teams with higher collaboration}; \textbf{separate Dev and Ops teams with a facilitator(s) in the middle}; \textbf{smaller Ops team with more responsibilities for the Dev team}; and \textbf{no visible Ops team}. Other authors also have attempted to describe one or several team structures and how their characteristics impact the software delivery performance\cite{humble2011enterprises,leite2021,lopez2021,luz2019,macarthy2020,sregoogledevops,nybom2016,stateofdevops,zhou2022cross}. All these empirical studies seek to understand and categorize team structures, also known as team taxonomies\footnote{Collections of classes wherein each class is an abstraction that describes a set of properties shared by the instances of the class~\cite{ralph:2019}}. They are representations of groups organized by their characteristics, such as \textbf{silos}, \textbf{communication}, \textbf{collaboration}, \textbf{automation} and others. 

Specifically, Leite et al. \cite{leite2021} identified four common organizational structures: (1) \textbf{Siloed Departments}, (2) \textbf{Classical DevOps}, (3)\textbf{ Cross-functional Teams}, and (4) \textbf{Platform Teams}. López-Fernández et al. \cite{lopez2021} also identified four organizational structures: (A) \textbf{Interdepartmental Dev \& Ops collaboration}, (B) \textbf{Interdepartmental Dev-Ops team}, (C) \textbf{Boosted cross-functional DevOps team}, and (D) \textbf{Full cross-functional DevOps team}. These models are not equivalent, but they have much in common. For example, (1) and (A) are practically the same team structure. Both are characterized by a certain collaboration of Dev and Ops teams but well-defined responsibilities and suffer silo-related problems and low performance. Nevertheless, (1) and (2) are not supported by any horizontal team, whereas (A) and (B) are usually supported by a platform team, usually limited to providing the required DevOps infrastructure.

Comprehending these theories and representations of team taxonomies is critical to guide  companies in a more systematic and structured DevOps adoption process.  In a previous work \cite{diaz2022}, we proposed a model that harmonizes existing team taxonomies in the way organizations structure their development and infrastructure/operation teams around the DevOps culture and practices. 




\subsection{Harmonizing DevOps Taxonomies: T3 - A Team Taxonomies Theory}
\label{sec:back2}

This section presents an overview of our Team Taxonomies Theory (T3) \cite{diaz2022}. This theory comprises the following constructs: \textbf{Team}, \textbf{Management}, \textbf{Culture}, \textbf{Automation}, \textbf{Platform}, \textbf{Silo}, \textbf{Collaboration}, and \textbf{Communication}. A set of attributes characterizes the Team construct, such as \textbf{autonomy}, \textbf{alignment of dev\&ops goals}, \textbf{responsibility/ownership sharing}, \textbf{skills/knowledge sharing}, \textbf{stack\&tools sharing}, and \textbf{cross-functionality}, among others. Teams collaborate and communicate with a particular frequency and quality, aiming to reduce silos and achieve a faster delivery flow. \textbf{Development, Operation, Product, and Horizontal Teams} are specializations of the Team construct.

Team interactions may evolve with technological and organizational maturity. For example, an organization could be structured by traditional Development and Operation teams, supported by a horizontal team named ``Bridge Team''. Other organizations could be structured by Product teams that assume both dev and ops responsibilities, supported by a horizontal team named ``Enabling Team''. We identified four main structures, they are:

\begin{itemize}
    \item \textbf{Bridge DevOps Team}: This team mainly helps and supports development and operation teams by deploying and hosting applications in the platforms they build, monitor, and support. The engineers of the Bridge DevOps teams are the DevOps practices facilitators; hence, they create, deploy, and manage both the infrastructure (environments) and the deployment (CI/CD) pipelines. They may also be involved in other tasks, such as requirement analysis and coding. They are usually the bridge interface between developers and IT Operations, driving the DevOps values and practices.

    \item \textbf{Enabling DevOps Team}: Organizations create specific teams to satisfy product team necessities. It includes platform servicing and tools (mainly for infrastructure and deployment pipelines), consulting, training, evangelization, mentoring, and human resources. Thus, they behave as enabler teams by providing these capabilities. Sometimes, it is unnecessary to create new teams because the operations team takes over these needs and assists product teams. The Enabling DevOps Teams are named in different ways, e.g., DevOps Centers of Excellence, chapters, guilds, platform/SRE teams, among others.
    
    \item \textbf{Product Team}: This Team is entirely responsible and has complete autonomy over a product or service (scoping, managing, architecting, building, and operating it). Therefore, it is common for it to have characteristics such as \textbf{responsibility/ownership sharing} about all the artifacts and tasks. In this way, product team can innovate quickly with a strong customer focus by aligning dev and ops objectives with business goals.
    
    \item \textbf{Development and Operations Teams}: These teams have well-defined and differentiated roles with their departments and objectives. The development team, for example, focuses on implementing features and is led by a project manager. The interaction between teams typically occurs as a work transfer. The developers deliver the finished code to the Operation team that creates and manage the infrastructure, while also responsible for the deployment.

\end{itemize}

\subsection{Other theories: Team Topologies' book}

The book \textit{Team Topologies: organizing business and technology teams for fast flow} by M. Skelton and M. Pais, provides insights into how software teams can achieve high performance, i.e., deliver value continuously and sustainably. Although this book does not present scientific evidence in building their theory, it is based on many experiences and real cases, becoming a well-accepted theory, widely used by practitioners \cite{leite2021organization}. Consequently, although we did not use it in the theory development phase, we now selected it to incorporate other perspectives in the research. To that end, Team Topologies proposes an adaptive model for organizational structure and team interactions, treating teams as the fundamental means of software delivery (the team is considered the primary unit of work to achieve shared agreed goals). 

The book presents three main concepts: cognitive load, team APIs, and team-sized architectures. Cognitive load is the amount of information and knowledge necessary to perform the work: \textit{“as team size goes up, complexity increases and so cognitive load goes up accordingly''}. Team APIs refer to any standard way to interact with the team, such as SLA, error budget, and documentation. Team-sized architectures is an approach to scale products and/or services based on the principles of end-to-end ownership and concerning the cognitive capacity of the team.

The book proposes four fundamental team topologies:

\begin{itemize}

\item \textbf{Stream-aligned team}: It is a cross-functional team aligned to a single work stream. It is empowered to build and deliver customer or user value as quickly, safely, and independently as possible without requiring handoffs to other teams to perform parts of the work. Thus, a stream-aligned team has minimal ---ideally zero--- handoffs of work to other teams, requiring a high degree of autonomy. This team structure is also known as a \textit{functional or traditional product application team}. 


\item \textbf{Complicated sub-system team}: It is a specialized team focused on an area of extreme complexity. They help avoid the extreme cognitive tax these complicated components would levy on stream-aligned teams (e.g., payment feature). 

\item \textbf{Enabling teams}: It is a team whose members work with other teams to develop missing capabilities.  Enabling teams focus on developing the other teams and guiding them toward the intrinsic motivators (autonomy, mastery, and purpose). 

\item \textbf{Platform team}: It is a team that provides stream-aligned teams with an internal software platform to reduce cognitive load.  \textit{“The purpose of a platform team is to enable stream-aligned teams to deliver work with substantial autonomy''}. An example is a cloud platform team that offers self-service infrastructure services.
\end{itemize}

Any type of team can interact with any other team through these three interaction modes:

\begin{itemize}
    \item \textbf{Collaboration}: It is ideal when a high degree of adaptability or discovery is needed to explore new technologies or techniques because it avoids the transfer of work between the teams. Collaboration occurs between teams with different skills and knowledge to bring together many people's combined knowledge and experience to solve challenging problems.
    \item \textbf{X-as-a-Service}:  It is suited to situations one or more teams use a code library, component, API, or platform, which can be effectively provided “as a service”. As the service is now available on a platform, it reduces the need for detail and context transfer across teams, thus limiting the cognitive load.
    \item \textbf{Facilitating}: It is suited for situations where one or more teams would benefit from the active help of another team facilitating or coaching some aspect of their work. This mode provides support and capabilities to many other teams, helping to enhance productivity or to solve some problems.
\end{itemize}

\section{Research Methodology} 
\label{sec:methodology}

\begin{figure}[ht]
\centering
\includegraphics [width=12cm]{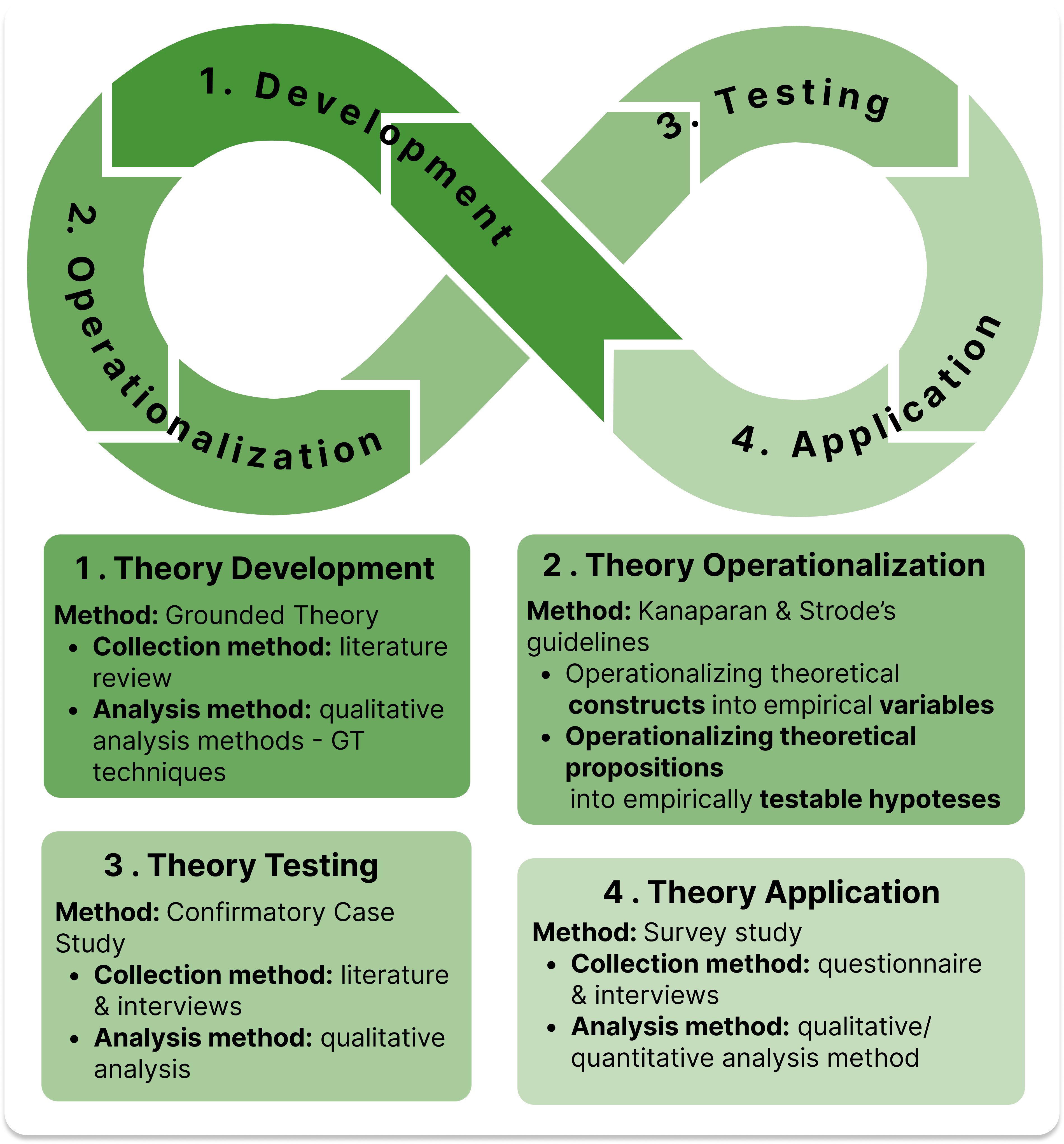}
\caption{Methodological framework.}
\label{fig:research-framework}
\end{figure}

The contribution described in this paper is a part of broader research in theory-building in the context of DevOps Team Taxonomies. To analyze and describe this phenomenon, we adopted a multimethod or mixed approach~\cite{creswell:2017}, in which two different research groups participated. The outcomes of theory-building research are enriched by building theory from multiple research perspectives and methods \cite{lynham2002}. The methodological framework depicted in Figure~\ref{fig:research-framework} illustrates the stages of this broader research. 


Theory-building is a continuous process of refinement and improvement that consists of four stages (\cite{lynham2002}, p. 229):

\textbf{1. Conceptual development} - the conception of pertinent constructs and relationships through inductive and abductive processes~\cite{sjoberg:2008}. The output of this phase is an explicit, informed, conceptual framework that often takes the form of a model (\cite{lynham2002}, p. 232). 

\textbf{2. Operationalization} -  the conceptual elements must be converted into observable
entities and measurable units that can be further inquired into and (dis-)confirmed \cite{lynham2002,sjoberg:2008}.

\textbf{3. Testing} - the examination of the validity through empirical studies to purposefully inform and intentionally confirm or disconfirm/reject the theoretical framework central to the theory \cite{lynham2002,sjoberg:2008}. 

\textbf{4. Application} -  the study, inquiry and understanding of the theory in action, i.e., in real situations in practical disciplines \cite{lynham2002}.

The development of the first version of the Team Taxonomies Theory (T3) was presented in a previous work \cite{diaz2022}. We followed the guidelines of Sj{\o}berg et al. \cite{sjoberg:2008}, who presented the unique software engineering theory-building framework described until the date of writing this paper. Based on a qualitatively oriented theory-building research method, we conducted a collaborative Grounded Theory study \cite{diaz2021GT}. We collected and analyzed data from the literature to develop a conceptual model composed of a set of constructs, propositions, explanations, and scope conditions of the theory.

This paper conducts Phases 2 and 3, i.e., the operationalization of our conceptual model and its testing. Performing these phases, we aim to increase the criteria for evaluating theories, named \textit{empirical support}, that is, \textit{“the degree to which a theory is supported by empirical studies that confirm its validity”} \cite{sjoberg:2008}.


\subsection{Theory Operationalization procedures}
\label{sec:method-operationalization}

The purpose of the operationalization phase is essentially \textit{“an explicit connection between the conceptualization phase and practice. The operationalization of a theory needs to be confirmed and/or tested in its real-world context”} (\cite{lynham2002}, p. 232). Thus, the operationalization addresses two aspects (\cite{sjoberg:2008}, p. 327):

\begin{itemize}
    \item Operationalizing theoretical constructs into empirical variables.
    \item Operationalizing theoretical propositions into empirically testable hypotheses.
\end{itemize}
	
Operationalizing the constructs involves three steps \cite{pritha2022}: (i) Identifying the concepts under study. (ii) Choosing one or more variables to represent each concept. For example, while it is relatively easy to assign variables to concepts, such as the variable $age$ to represent the length of time that a person has lived, this is not the case for concepts such as anxiety or quality of life. Finally, (iii) to select indicators for each variable, i.e., define the range of values (type) they can take at different times. These indicators must have a reasonable level of reliability. For example, the variable $frequency$ could represent the concept $collaboration$ and one could assign the indicators $daily$ and $eventual$; however, different researchers may give different semantics to the value $eventual$, which makes the experiment non-reproducible. 

Operationalizing the propositions mainly consist in establishing testable hypothesis from propositions. Propositions have to be converted into hypotheses involving the variables and indicators previously defined. In the words of E. Indrajit \cite{eko2006}: \textit{“A hypothesis is the predictions about the values of the units of a theory in which empirical indicators are employed for the named units in each proposition”}. Kanaparan and Stroke \cite{Kanaparan:2021} provide six detailed guidelines for \textit{the transition from theory building to theory testing} with a focus on the activities and reasoning. It occurs during the development of a testable research model; i.e. they indicate how to move from propositions to hypotheses. An exemplar of theory operationalization is described by Luna et al. \cite{luna2020}, in which the authors presented a survey on agile governance and identified 11 propositions, 24 indicators, and 16 hypotheses.

\subsection{Theory Testing procedures}
\label{sec:method-testing}

Testing is the last step of the theory-building process, which involves \textit{“examination of the validity of the theory’s predictions through empirical studies”} (\cite{sjoberg:2008}, p. 327). Testing consists in contrasting the hypotheses obtained from the operationalization process with the information generated through the analysis of case studies. These confirmatory case studies are especially important for disproving theories. \textit{“The detailed insights obtained from confirmatory case studies can also be useful for choosing between rival theories”} (\cite{sjoberg:2008}, p. 296). Note that \textit{“each study adds more evidence for or against the propositions of the theory”} (\cite{sjoberg:2008}, p. 293). Thus, case studies constitute an instrument to prove the theoretical framework's adequacy.

The method we selected to evaluate the theory is a \textbf{confirmatory case study}. We collected data from the literature, specifically the book “Team topologies: organizing business and technology teams for fast flow” by M. Skelton and M. Pais and interview with M. Pais. The data gathered from those sources were stored in a tool for qualitative data analysis and research, ATLAS.ti \cite{Atlas:2019} and are available in our public repository. To analyze the data, we conducted initial/open coding and selection of core categories in several iterations. The outputs of these phases are the codes, the memos, and the categories, all managed through the Atlas.ti v9 tool \cite{Atlas:2019}. To support our findings, we also included some excerpts from the collected data to support the chain of evidence.

\section{Operationalizing T3}
\label{sec:theory-operationalization}

The operationalization of a theory addresses two activities: i) operationalizing theoretical constructs into empirical variables; and ii) operationalizing theoretical propositions into empirically testable hypotheses (Section~\ref{sec:method-operationalization}). This section describes how we conducted both activities to operationalize T3.

\subsection{Operationalizing theoretical constructs into empirical variables}
\label{sec:construct-operationalization}

Table~\ref{tab:constructs} describes the constructs of T3 (see \cite{diaz2022} for a full description of constructs). It also shows the variables representing the constructs and the indicators (type) for each. The semantics of the indicators in Table~\ref{tab:constructs}  may be open to subjective interpretation. Thus, the semantics of each of the indicators are refined and detailed in Table~\ref{tab:indicators}. When necessary, some excerpts from \cite{diaz2022} were provided (in quotes) to facilitate understanding the indicators.

\begin{longtable}{p{5cm}p{4cm} p{3cm}}
\caption{Constructs, variables, and indicators. If the indicator is expressed in brackets, the variable can take only one of the values expressed in brackets. If the indicator is expressed in square brackets, the variable can take several values of those expressed in square brackets.}
\label{tab:constructs}
\\\hline\textbf{ Concepts (constructors)} & \textbf{Variables} & \textbf{Indicators} 
\endhead
\hline
\hline 
\multirow{17}{*}{   \begin{tabular}[c]{@{}p{5cm}@{}} C1 to C3. Team. It is   an artificial (abstract) concept that represents a team structure of an IT   department \end{tabular} } & autonomy & (dependent,   self\_organization)   \\ 
 & & \\
 & blame & (true, false)   \\
 & & \\
 & alignment\_of\_dev\&ops\_goals & \begin{tabular}[c]{@{}p{3cm}@{}} (local\_optimization, \\ product\_thinking) \end{tabular}  \\
 & & \\
 & responsibility/ownership\_\-sharing & (full\_sharing,   medium\_sharing, minimal\_or\_null\_sharing)   \\
 & & \\
 & skills/knowledge\_sharing & (full\_sharing,   medium\_sharing, minimal\_or\_null\_sharing)   \\
 & & \\
 & stack\&tools\_sharing & (full\_sharing,   medium\_sharing, minimal\_or\_null\_sharing)   \\
 & & \\
 & cross-functionality/skills & (true, false)   \\
 & & \\
 & role\_definition/attributions & (true, false)  \\
 & & \\
 & inheretedMembers & (product teams,   horizontal teams, bridge teams, enabler teams, development teams, operation   teams)  \\ \hline
 
 & & \\
\multirow{5}{*}{    \begin{tabular}[c]{@{}p{5cm}@{}}C4. Management.   It is an artificial (abstract) concept that represents the management   activities (project and product management, change management, team   self-organization, coordination and transfer of work between teams, as well   as measuring, monitoring and feedback) to ensure that software development   and maintenance are systematic, organized, and quantified. \end{tabular} } 
& change\_type & (small\&frequent,   large\&rare)  \\
 & & \\
 & product\_name & String   \\
 & & \\
 & inheritedMembers & \begin{tabular}[c]{@{}p{3cm}@{}}(project\_management, \\  product\_management) \end{tabular}  
  \\  \\\\ \\\\\\\hline
 
\multirow{11}{*}{    \begin{tabular}[c]{@{}p{5cm}@{}}C5. Culture. It   represents the set of common principles, values, and best practices that   impact on the way that people in an organization relate to each other and   make decisions and actions around application life-cycle management. \end{tabular}} & principles & {[}customer\_centric\_action,   end\_to\_end\_responsibility, automate\_every\_thing{]}  \\
 & & \\
 & cultural\_values & {[}communication,   collaboration, transparency, blame{]}   \\
 & & \\
 & best\_practices & \begin{tabular}[c]{@{}p{3cm}@{}} {[}continuous\_integration, \\  continuous\_testing, \\ continuous\_delivery, \\continuous\_deployment, \\ continuous\_monitoring, \\ continuous\_improvement, \\ infrastructure\_as\_code{]}  \end{tabular} \\ \hline
 
  & & \\
  
  \begin{tabular}[c]{@{}p{5cm}@{}} C6. Automation. It represents the set of activities to automate the processes of the application life-cycle management and infrastructure management where applications are deployed. \end{tabular} & type & (Automated\_Infras\-tructure\_Management, Automated\_Application\_Li\-fe\_Cycle\_Management)  \\ \hline
 & & \\

  \begin{tabular}[c]{@{}p{5cm}@{}} C7. Platform. It represents the technology to support automation. It provides interfaces for \textbf{Automated infrastructure management} and automated application life-cycle management. \end{tabular} 
  & provided\_interface  & (IaC,  ALM\_Interface),(Auto\-mated\_Infrastructure\_Ma\-nagement, Automated\_Application\_ Life\_Cycle\_Management)  \\ \hline

 & & \\
\begin{tabular}[c]{@{}p{5cm}@{}} C8. Silo. It   represents the concept of silos between teams. The organizational silo represents the teams that are organized in different locations. Cultural silos are remnants of organizational silos, where teams work together. However, there is still a work transfer culture, poor interaction, and communication between teams, promoting work transfer. \end{tabular} & type & (organizational, cultural)  \\ \hline

 & & \\
\begin{tabular}[c]{@{}p{5cm}@{}}C9. Collaboration between teams. From the lack or eventual
collaboration, to daily collaboration \end{tabular} & frequency & (daily, eventual)   \\ 
 & & \\
 & quality & (high, low)  \\ \hline
 
  & & \\
\begin{tabular}[c]{@{}p{5cm}@{}}C10. Communication between teams. From poor/rare communication to
frequent communication. \end{tabular}  & type & (poor/rare,   frequent) \\ \hline
\end{longtable}

\begin{table}[]
\caption{Refinement of the meaning of the main indicators.}
\label{tab:indicators}
\begin{tabular}{p{6,2cm} p{6cm}}
\hline \textbf{Variable: Indicator }& \textbf{Definition}   \\ \hline
autonomy: self\_organization & A team can access to all necessary information to develop, deploy, and operate the product. “Now   developers have autonomy for going from zero to production without having to   wait for anyone”. \\
autonomy: dependent & A team depends on another team/party at some point in the product lifecycle. \\
blame & It is part of the team culture values. It occurs when the Dev or Ops team blames the other for a problem or error, usually occurs when they don't share the responsibility. \\
\hline
alignment\_of\_dev\&ops\_goals: product\_thinking & Dev and ops goals are aligned towards the product. “A collaborative culture requires product thinking, in substitution to operations or development thinking.” \\
alignment\_of\_dev\&ops\_goals: local\_optimization & Dev and ops have their own goals. \\
\hline
responsibility/ownership\_sharing: full\_sharing & Dev and ops share responsibility of products, output artifacts (e.g., databases), and tasks (e.g., NFR shared responsibility, infrastructure management shared responsibility, monitoring shared responsibility, and incident handling shared responsibility, etc.) so that everyone is responsible for build, test, deploy, operate and maintain their applications and infrastructure. \\
responsibility/ownership\_sharing: medium\_sharing &   Dev and ops may share some responsibility of products, output artifacts (e.g., databases), and tasks (e.g., NFR shared responsibility, infrastructure management shared responsibility, monitoring shared responsibility, and incident handling shared responsibility, etc.) although some responsibilities may still reside on one side (dev or ops). \\
responsibility/ownership\_sharing: null\_sharing & Dev and ops have separate responsibilities and tasks (each team member has different responsibilities and tasks). \\
\hline
skills/knowledge\_sharing: full\_sharing & High levels of knowledge sharing (e.g., developers may have knowledge about infrastructure/platform). \\
skills/knowledge\_sharing: medium\_sharing & Medium levels of knowledge sharing or minimal awareness of what is happening on the other side of the wall. \\
skills/knowledge\_sharing: null\_sharing & Low levels of knowledge sharing or no awareness of what is happening on the other side of the wall.\\
\hline
stack\&tools\_sharing: full\_sharing & A holistic view of the tools and stack—--the frontend, backend, libraries, storage, kernels, and physical machine, which have a high degree of standardization and integration. \\
stack \&tools\_sharing: medium\_sharing &  Partial view of the tools and stack. Not all the tools have been standardized and integrated, such monitoring tools.\\
stack \&tools\_sharing: null\_sharing & Non-shared stack. \\
\hline
cross-functionality/skills & An attribute of teams that have the necessary skills to achieve end-to-end ownership. \\
\hline
role\_definition/attributions & When true, it indicates when a team has well-defined roles and are less likely to share responsibility. When false, it indicates that the teams do not have a well-defined definition of responsibility. \\
\hline
change\_type: small\&frequent &  With a small batch size (i.e. the unit of work that passes from one stage to the next stage in a process), each batch makes it through the full lifecycle quicker, at least once a day according to Fowler definition for continuous integration.\\
change\_type: large\&rare &  Large batch size hampers continuous integration and delivery. “Pushing 2 weeks of commits to prod is a larger batch than pushing every commit in a continuous delivery system”.\\
\hline
collaboration\_frequency: eventual/daily  & The frequency of collaboration between teams when they share responsibility for an activity.  \\
collaboration\_quality: high/low &  It represents the quality of collaboration between teams. \\
\hline
communication\_type: poor/rare   &  \\
communication\_type: frequent   & It presents the level of communication established between the teams, which can range from rare to frequent.\\
\hline
\end{tabular}%

\end{table}

\subsection{Operationalizing theoretical propositions into empirically testable hypotheses}
\label{sec:proposition-operationalization}

Among the different types of hypotheses (null, alternative, statistical, simple, and compound), we selected $causal$ hypothesis as a means of expressing empirically testable hypotheses. This type of hypothesis establishes a cause-effect relationship between independent and dependent variables. When necessary, we also used relational hypotheses that only express the existence of a relationship between variables without asserting anything about their variation.

Kanaparan and Strode \cite{Kanaparan:2021} stated that: \textit{“By identifying the independent and dependent variables, the research problem is then presented in a form that enables the presentation and evaluation of a cause and effect relationship”}. Therefore, the first step is establishing which variables are dependent and which are independent~\cite{Kanaparan:2021}. The causal hypothesis states that a change in the independent variable will generate changes in the dependent variable. We specified this set of causal relationships. The labels \textit{independent} and \textit{dependent variables} only establish a role, i.e. the variation of the variables expressed in the columns can affect the variables stated in the rows. The list of variables, their classification into dependent, independent variables and their correlations are depicted in Appendix \ref{app:propositions}.
The following two hypotheses exemplify the output of the operationalization process, highlighting both variables and values of the indicators:

\textbf{H1.} A team culture based on \textbf{responsibility/ownership sharing} \textit{enables} \textbf{daily collaboration} between team members

\textbf{H2.} A team culture based on \textbf{responsibility/ownership sharing} \textit{enables} \textbf{high quality collaboration} between team members

For all other relationships (i.e., \textit{reduce, impact, require}, and \textit{increase}), the hypotheses are constructed similarly, except for the $is\_a$ relationship, where the hypothesis does not determine a cause-effect relationship but only a type of association (relational hypothesis). The hypotheses generated are provided in Appendix \ref{app:propositions}.

\section{Confirmatory case study}
\label{sec:usecase}

This section describes the case study that provides empirical evidence that confirms or refutes T3. Case study research is a technique that investigates contemporary phenomena in their natural context to search for evidence, gain understanding, or test theories by primarily using qualitative analysis \cite{runeson09}. Our objective is to test our theory T3. Thus, we designed a confirmatory case study after operationalizing T3 (see Sections~\ref{sec:construct-operationalization} and \ref{sec:proposition-operationalization}).

The nature of our theory makes it a candidate to seek a case study broad enough to analyze the reality of companies with several organizational and team structures. Hence, we concluded that collecting data from a case study in the context of a single company would have been less enriching, since it would not likely have all team structes covered in T3.

We collected data from an existing theory named \textit{DevOps Topologies}. This theory has been expanded beyond DevOps alone in the \textit{Team topologies: organizing business and technology teams for fast flow} book by M. Skelton and M. Pais \cite{skelton2019team}. We did not use the book as a source during the theory development. We selected it to the testing phase because the book describes several team structures and has become a reference for practitioners. Additionally, we interviewed the book author and DevOps consultant Manuel Pais. He is a consultant whose focus is to help organizations build a DevOps culture, automate processes to accelerate value delivery, and understand technical and non-technical aspects of DevOps, specifically those related to organizational team structures. Therefore, M. Pais is familiar with many companies with different organizational structures. Collecting and analyzing these data allow us to verify whether our theory is in accordance with DevOps Topologies and the accumulated knowledge of M. Pais and M. Skelton.

We conducted this case study inspired by the methods presented by Robert K. Yin~\cite{yin:2018} for planning, designing, collecting, and analyzing the different fields and ways of applying them. In addition to the data collected from the book, we designed a semi-structured interview protocol  to validate or refute these hypotheses. 
We organized the questions into five blocks (those related to collaboration, responsibility/ownership sharing, cross-functionality/skills, automation, and platform/enabling teams). At the beginning of each block, we elaborated on a generic question to introduce the context we wanted to explore with the interviewee. For example, (\textit{ Block 1. We understand that collaboration is crucial in organizing an IT department's teams. What factors promote collaboration, or what aspects are impacted by more or less collaboration?}).  Then, for each T3 proposition, we had at least one related question. For instance, seeking to validate H1.1 \textit{A team culture based on responsibility/ownership sharing enables, at least, an eventual collaboration between teams}, we had this question: \textit{The importance of a culture of shared product ownership or product responsibility is widely known. In what particular aspects does this shared ownership/responsibility impact (for example, it could be said that it promotes collaboration)?}. 
The initial overview, interview protocol, and transcription are available in the supplementary material.

From the book, we selected Chapters 5, 6, 7, and 8. We exclude Chapters 1 to 3 because they explain the definition of team topology and discuss what cognitive load and Conway's law are. This information is essential to understand the concepts necessary to describe a team structure. However, it is not helpful to validate or refute our theory. On the other hand, Chapters 5,6,7, and 8  present concepts such as team topologies/structures, interaction modes, boundaries, and cognitive load related to structures wich are important in our context.

We analyzed the documents (the book chapters and the interview transcript) inspired by the constructivist GT. In this process, four raters (coders) were involved: Isaque Alves; Jessica D\'iaz; Daniel L\'opez-Fern\'andez; and Carla Rocha. Initially, we generated some codes using open coding, which aims to discover the concepts underlying the data and instantiate them in the form of codes. Some codes were related to the constructs and propositions of our theory T3  (e.g., \textit{Collaboration requires alignment and sharing of responsibilities}). Other codes were not mentioned in our theory (e.g., \textit{Enabling teams reduce cognitive load}), and we discuss further in Section \ref{sec:discussion}. 
The final codebook consists of $111$ codes, organized into $14$ categories (e.g., \textit{Automation, Team Attributes, and Team Interactions Patterns}),  the memos, and excerpts from the interview and chapters. These artifacts are the input for the testing the theory.

\section{Hypotheses Testing}
\label{sec:hypoheses-validation}

In this section, we test our theory through the confirmatory case study described above. We confront the hypotheses developed in the operationalization process with the codes generated with the case study analysis. The outputs of this qualitative analysis (i.e., the codebook with categories and codes and the memos) are used for contrasting the hypotheses. To support our findings, we also included excerpts from the collected data to maintain the chain of evidence. Excerpts are identified with the document ID and quotation ID, just like Atlas.ti manages quotations (e.g., [1:2] means Quotation 2 in Document 1). 

\subsection{Hypotheses about Collaboration}

According to our theory, \textbf{Collaboration} is one of the most important constructs of the team taxonomy, as it deals with the different forms of relationships between dev and ops. \textbf{Collaboration} between teams has two attributes: frequency (eventual and daily) and quality (high and low). 

First, we check whether the construct \textbf{Collaboration} is in the collected data. In this regard, we found \textbf{Collaboration} is defined as a \textit{Team Interaction Mode} together with other interaction modes such as $Facilitating$ and $X\_as\_Service$ (groundedness\footnote{Groundedness is the number of quotations labeled with a code} = 88). \textbf{Collaboration} is described in the data as [2:12] \textit{“working closely together with another team''}. A memo explains that \textit{“Even if there are Dev \& Ops silos, collaboration must be well defined''}. 

We also revisited the hypotheses related to collaboration, seeking to support or refute our theory.

\vspace{0.2cm}
\noindent \fbox{\begin{minipage}{12cm}
H1.1 A team culture based on \textbf{responsibility/ownership sharing} \textit{enables}, at least, eventual \textbf{collaboration} between teams. 

H1.2 A team culture based on \textbf{responsibility/ownership sharing} \textit{enables} \textbf{high quality collaboration} between teams.
\end{minipage}}
\vspace{0.2cm}

Second, we analyzed the collected data to determine if we could find evidence of whether responsibility/ownership sharing and \textbf{collaboration} are related. The codes \textit{“Collaborating creates a blurring of responsibility boundaries that leads to responsibility sharing''} and \textit{“Collaboration implies alignment and sharing of responsibilities''} corroborate with the hypotheses. The following excerpt supports these codes [2:2] “\textit{...the two teams must take on joint responsibility for the overall outcomes of their collaboration because the act of collaborating creates a blurring of responsibility boundaries. Without joint responsibility, there is a danger of losing trust if something goes wrong.}".

Third, we searched whether the collected data provided more evidence about the \textbf{frequency and quality of collaboration}. In the interview, M. Pais presented an example of Dev \& Ops teams that started a collaboration. [3:56] \textit{“... collaboration is quite frequent at the beginning, and then there will be less collaboration because the development team can do certain deployments themselves or create some infrastructure. Over time you may notice that you do not have to collaborate weekly, but monthly [...]''}. The collected data also highlighted that the quality of collaboration has more value than frequency. The following excerpt support this argument:  [3:50] \textit{“More than with a specific frequency (daily, weekly, monthly), it is that the collaboration is dedicated and of quality. It is better to collaborate once a month, but we have been working together for a week and are not trying to do other things''}. Consequently, hypothesis H1.1 and H1.2 are supported by the case study, as responsibility/ownership sharing and collaboration (at least occasional and highquality collaboration) are two inter-related concepts, in such a way that collaboration leads to responsibility/ownership sharing and vice-versa responsibility/ownership
sharing enables collaboration.

\vspace{0.2cm}
\noindent \fbox{\begin{minipage}{12cm}
H2.1 \textbf{High quality, day-to-day collaboration} \textit{reduces} \textbf{organizational silos}.

H2.2 \textbf{High quality, day-to-day collaboration} \textit{reduces} \textbf{cultural silos}.
\end{minipage}}
\vspace{0.2cm}

For the collaboration quality hypotheses, we searched in the collected data whether collaboration and the existence (or non-existence) of organizational and cultural silos are related.
When there are silos in an IT department, teams have no knowledge about each other's work, i.e., the development team does not know how to create infrastructure or how to monitor the application/system when it is deployed  [3:49] “\textit{...at this point, teams need much collaboration}". Collected data also shows that  [3:50] “\textit{We understand that collaboration is not only about executing a task but also about starting to reduce silos. To that end, collaboration should be dedicated and of quality. You have to give it value and allow the teams to be $100\%$ focused on that collaboration}". 
 
Therefore, hypotheses H2.1 and H2.2 are partially supported. The data emphasize the role of collaboration quality in reducing silos (without distinction between cultural or organizational) but not on the direct impact of the frequency of such collaborations. 

\vspace{0.2cm}
\noindent \fbox{\begin{minipage}{12cm}
H7.1 \textbf{collaboration} requires that dev \& ops teams share business goals and global ones (product thinking).
\end{minipage}}
\vspace{0.2cm}

We analyzed the collected data to check whether collaboration is related to the alignment of Dev \& Ops goals among team members. The collected data showed a typical situation of teams and departments with poorly defined objectives and conflicting incentives, putting one against the other. The data highlighted that blaming culture needs to be solved before the teams can collaborate. If these teams need to collaborate, existing conflicts will tend to get worse [3:6] “\textit{if your company's incentives are putting one team against another, then you have a problem to solve. Putting together people from different teams who are used to blaming each other only makes the situation worse.}". Therefore, teams must have aligned goals and incentives for collaboration to result in innovation and problem-solving. 
Consequently, hypothesis H7.1 is supported by the case study.

\subsection{Hypotheses about sharing of responsibility/ownership}

\textbf{Sharing responsibility/ownership} is a key attribute of the Team construct. This variable may assume the following values: $full\_sharing, medium\_sharing$, and $null\_sharing$. 

We checked whether this variable is in the collected data. We found that sharing responsibility/ownership is mentioned several times as a fundamental team attribute (groundedness = 15).

Once again, at this stage, we analyze the hypotheses related to sharing of responsibility/ownership, looking to support or refute our theory.


\vspace{0.2cm}
\noindent \fbox{\begin{minipage}{12cm}
H9.1 \textbf{Cross-functional} teams are \textit{characterized by} sharing all or some responsibility for products, output artifacts, and tasks so that everyone is responsible for building, testing, deploying, operating and maintaining their applications and infrastructure.
\end{minipage}}
\vspace{0.2cm}

To confirm or refute hypothesis H9.1, we searched whether \textbf{responsibility/ownership sharing} is a characteristic of \textbf{Cross-functional} teams in the collected data. Looking at the code category \textit{Team Attributes}, we found the \textit{“Cross functional"} code and the \textit{“Shared ownership and end-to-end vision"} code. The following excerpt [1:4] supports these codes \textit{“[...] teams must be cross-functional and include all the required capabilities to manage, specify, design, develop, test, and operate their services"}. Also, a member of a cross-functionality team must be able to assume different roles (e.g., develop, test, and deploy). The mentioned codes also support this idea. Indeed, the collected data shows that [1:4] \textit{“...each individual has a primary area of expertise, but their contributions are not limited to it"} and mentioned during the interview [3:73] the T-shape and Comb-shape models, in which \textit{“each person has a speciality, but he/she can do many different things"}. 

Last, regarding the hypothesis under study and the mentioned codes, M. Pais warned us about an anti-pattern that sometimes occurs in companies due to a misunderstanding of the cross-functional concept. Specifically, he claimed that [3:73] \textit{“...having a 15-people product team with a designer, a frontend, a backend, a tester, a PO, etc in which they don't have shared responsibility because each one is in his silo within the team is a bit naive."} and concluded that \textit{“...it is better to keep a small team and go for cross-functionality with more shared responsibility"}.

Therefore, hypothesis H9.1 is fully supported by the collected data.

\vspace{0.2cm}
\noindent \fbox{\begin{minipage}{12cm}
H11.1 The existence of \textbf{enabler (platform) teams} becomes facilitators and makes \textbf{ownership sharing} possible.
\end{minipage}}
\vspace{0.2cm}

In the code category \textit{Platform teams attributes}, we found the codes \textit{“Platform teams help to promote shared ownership in product teams"} and  \textit{“Platform teams reduce the cognitive load of teams that consume the platform"}, which are related to our hypothesis. Several excerpts support these codes. For example, the following [1:10]: \textit{“The purpose of a platform team is to enable stream-aligned teams to deliver work with substantial autonomy. The stream-aligned team maintains full ownership of building, running, and fixing their application in production"}. In this way, the usage of effective platforms reduces the cognitive load of specific tasks and makes it easier, for example, for a software developer to deploy his/her software with relative ease, thus enabling the T-shape model mentioned above.

Therefore, hypothesis H11.1 is fully supported by the collected data.

\vspace{0.2cm}
\noindent \fbox{\begin{minipage}{12cm}
H12.1 If a team in the autonomy attribute has the value self-organisation, this implies that there is full/medium responsibility/ownership sharing.
\end{minipage}}
\vspace{0.2cm}

To confirm or refute this hypothesis, we found the \textit{“Autonomy requires shared ownership"} code. Regarding this code, M. Pais stated that [3:39] \textit{“An autonomous team has to share more responsibility within the team"} and went further by saying that \textit{“these responsibilities to be shared also include product definition and customer value definition"}. Furthermore, in the collected data, we found the code \textit{“Shared ownership is promoted through personal intrinsic motivators: purpose, mastery, and autonomy"}. This code is supported by the following excerpt [3:59]: \textit{“...to have autonomy to take responsibility for one's work is a key factor for team members to be motivated to share responsibility"}.

Therefore, autonomy and responsibility/ownership sharing are related. Thus, H12.1 is fully supported by the collected data.

\subsection{Hypotheses about cross-functionality/skills}
According to our theory, cross-functionality/skills is a key attribute of the Team construct. 
First, we checked whether this variable is in the collected data. Cross-functionality is mentioned several times as a fundamental team attribute (groundedness = 10).

\vspace{0.2cm}
\noindent \fbox{\begin{minipage}{12cm}
H18.1 \textbf{Enabler teams} need a variety of skills to develop a \textbf{platform}. 
\end{minipage}}
\vspace{0.2cm} 

From the collected data, hypothesis H18.1 is supported. A platform team, [3:80] ``\textit{…these platform teams provide an internal service or product to the other teams… You have to understand the product team's use case. All this is the same as a stream-aligned team outside or inside the platform''}. 

\vspace{0.2cm}
\noindent \fbox{\begin{minipage}{12cm}
H19.1 \textbf{Multidisciplinary/poly-skilled} teams (i.e., teams with all the necessary skills such as development, infrastructure, etc.) avoid organizational silos.
\end{minipage}}
\vspace{0.2cm} 

From the collected data, hypothesis H19.1 is also supported. Product teams (aka. stream-aligned teams according to the Team Topologies' theory) must own cross-functional skills. [3:80] ``\textit{And they should also have cross-functional skills, everyone should know how to design (even if it is an API, it has design too, it can be easy or difficult to use)''}.

\subsection{Hypothesis about automated application life-cycle management}
According to our theory, automated application life-cycle management is a key construct, which is mainly provided by a platform, with  code category \textit{Automation} (groundedness = 17).

\vspace{0.2cm}
\noindent \fbox{\begin{minipage}{12cm}
H21.1 \textbf{organizational silos/conflicts} make the adoption of an \textbf{automated application life-cycle management} difficult
\end{minipage}}
\vspace{0.2cm}

Hypothesis H21.1 is supported. The silos [3:92] ``\textit{...makes the automation very difficult because instead of automating it, each silo tries to do it within its silo. This takes a lot of time or effort, or they try to use services or even shadow projects that are not official and transparent. This is a cultural problem since they cannot openly discuss their problems}''. The collected data also showed that [3:92] ``\textit{In general, silos make it difficult. In other words, it is not 100\% bad because there are cases where you want to automate what this team or this group of teams needs in a specific way. After all, it is what helps}''. [3:93] ``\textit{A typical example is that a development team uses Jira to manage their work and the operations team uses something more specific like ServiceNow. The problem is that if you have a solid silo, there is no connection between these two related tools. For example, if you are making a change in Jira, in the end, you have to do a deployment, and then you have to open a ticket in ServiceNow. However, the operation team would be interested in certain data from Jira because if there is a problem, they have to see what has changed in the code, who has requested this''}.

\subsection{Hypotheses about enabler platform teams}

According to our theory, enabler platform teams are an inherited element of the Team construct, mapped in the code category \textit{Enabling teams} (groundedness = 36). The role of this team is key since it helps product teams to learn 
new skills, thus helping them to reach the \textit{``you build it, you run it''} capability.

\vspace{0.2cm}
\noindent \fbox{\begin{minipage}{12cm}
H24.1 An \textbf{enabling team} allows product teams for autonomy: \textbf{self-organization} (can access all the information needed to develop, deploy and operate the product) without having to wait for anyone else. 
\end{minipage}}
\vspace{0.2cm}

To test hypothesis H24.1, we looked for information about the purpose of the enabling teams in the collected data. We found the code \textit{"Enabling team :: purpose"}, which contains several excerpts explaining the purpose and characteristics of this team structure. For example, this excerpt [1:2] explains the purpose \textit{``The mission of an enabling team is to help stream-aligned teams acquire missing capabilities, taking on the effort of research and trials, and setting up successful practices''}, meanwhile this excerpt [3:98] deepens the reflection \textit{``Enabling teams are small teams of people with strong specializations in an information domain who also work as mentors or teachers in a very active way with stream-aligned teams. An enabling team should be side by side with the stream-aligned teams helping them solve their specific problems for a long time''}. Although, the collected data explicitly enables teams to increase the product teams' autonomy. We can deduce that it happens since enabling teams teach product teams and help them to reach the \textit{``you build it, you run it''} capability. Therefore, H24.1 is supported in the collected data.

\vspace{0.2cm}
\noindent \fbox{\begin{minipage}{12cm}
H25.1 \textbf{Enabler team} provides \textbf{platform servicing} to product teams to assist them on DevOps platform.
\end{minipage}}
\vspace{0.2cm}

Regarding this hypothesis, it should be highlighted that Team Topologies' book organizes the responsibilities of horizontal teams differently from T3. In T3, enabling teams provide both platform servicing and consultancy. However, in Team Topologies' theory, platform teams take responsibility only for the platform servicing while enabling teams to take responsibility exclusively for the consultancy. The \textit{``Responsibilities of enabling teams can be incorporated into the platform team''} code has the following excerpt [3:102] \textit{``In many companies, the platform teams also do the facilitation''}, which would correspond to the model proposed in our theory. Therefore, the collected data partially support the presented hypothesis (H25.1).

\vspace{0.2cm}
\noindent \fbox{\begin{minipage}{12cm}
H28.1 Enabler (platform) teams provide automated application life-cycle management which is nothing more than a special case of platform servicing.
\end{minipage}}
\vspace{0.2cm}

The concern mentioned in the previous hypothesis also applied to the present one (H28.1) as the definition of horizontal teams is different in both theories. In the \textit{``Enabling team :: purpose''} code we can see that the contribution of enabler platform teams, as understood by our theory, provides automated application life-cycle management. This is evidenced by the following excerpt [3:102]: \textit{``Combined efforts of the platform and enabling teams is sufficient to achieve automated application life-cycle management''}. Since in T3, enabler platform teams assume the responsibility of platform and enabling teams (as understood by the Team Topologies theory), we can have that H28.1 is supported by the collected data.

\section{Discussion}
\label{sec:discussion}

Based on the results obtained, we now draw three discussions that we believe are of interest for this research.

\subsection{Theory Testing evaluation}




In this paper, we operationalized T3 by generating 34 testable hypotheses. From the interview and book analysis, we obtained 14 categories and 111 codes used to confirm/contrast each hypothesis. 

Of the 34 hypotheses, we tested 14, of which 11 were completely supported by the data and three were partially supported, reaching around 40\% test coverage. In this way, we tested the core of our theory, finding  great groundedness in the data to support the hypotheses. 



Among the testable hypotheses, 20 were not tested because we did not discuss these topics during the interview and we did not find evidence in the book. This can be explained because the interview protocol and the hypotheses were independently generated by different researchers to avoid bias, so that the researchers responsible from the interview protocol did not have access to the hypotheses. In any case, this is a first round of testing, in which we pursued testing the core of our theory. Successive data analysis rounds can focus on the not-tested hypotheses .

\subsection{Implications of results for practitioners}

Based on the main results of the T3 testing, a company focused on organizing and structuring the teams of its IT department has to focus on:

\begin{itemize}

\item Promoting collaboration in terms of quality and frequency, as long as such collaboration aims to reduce \textbf{organizational and cultural silos} and \textbf{align dev \& ops goals with business goals} in order to accelerate the flow of value delivery. 

\item Promoting \textbf{cross-functional teams} to reduce dependencies with other teams and avoid \textbf{organizational silos}. In turn, cross-functional teams promote \textbf{responsibility and ownership sharing}, so that everyone is responsible for building, testing, deploying, operating, and maintaining applications and infrastructure, which, in turn, makes collaboration easier. 

\item Promoting team \textbf{autonomy and self-organizing}. Autonomy enables  teams to share responsibilities on the different tasks for building, testing, deploying, operating, and maintaining applications and infrastructure.  

\item Creating \textbf{enabler teams} as facilitators (e.g. consulting, training, evangelization, and mentoring), which allows product teams for \textbf{autonomy} and \textbf{ownership sharing}. 

\item Creating \textbf{enabler teams} as \textbf{platform as a service} providers (e.g. platform for automated application life-cycle management), which allows product teams for \textbf{autonomy} and \textbf{self-organization}, that is, everyone can access all the information needed to develop, deploy and operate the product without having to wait for anyone else. This enabler teams require a variety of skills to develop the platform. 

\end{itemize}

\subsection{Suggested areas for improvement}

Beyond the hypotheses testing, the interview with M. Pais was really useful to understand that the organization and structure of teams in an company is not static. Although organizational charts are useful for some reasons, the day-to-day running of organizations requires continuous adjustments in their team structures or topologies by taking into account various socio-technical considerations. Some of these consideration are about Conway's laws\footnote{Any organization that designs a system will inevitably produce a design whose structure is a copy of the organization’s communication structure.}, Dunbar's number\footnote{A suggested cognitive limit to the number of people with whom one can maintain stable social relationships, which Conway applied to work relationships, describing an ideal team size of 5 people (Dunbar level 1), and possibly team size 15 people (Dunbar level 2).}, and the cognitive load\footnote{The idea that a software engineer has to spend mental effort for being able to safely make a change in a software system \cite{skelton2019team}.} and its relation with DevOps cross-functional teams with end-to-end responsibility and vision.  

Moreover, to think that we can put into practice the idea that ``everyone should collaborate with everyone'' that DevOps sometimes promotes is, in Pais' words, a bit na\"{\i}ve. It is clear that collaboration improves and solves problems involving different teams, driving innovation and rapid discovery. However, collaboration is expensive, reducing the work capacity of teams and increasing their work in progress. For M. Pais, collaboration becomes an investment, so it must be well-adopted. Due to the lack of maturity, according to M. Pais, many companies prefer to avoid teams collaboration because collaboration is expensive and its results are not immediate---compared to practices such as automation (e.g., CI/CD) in which the return of investment is more immediate. For the collaboration to have a return on investment, the collaboration must first specify the teams involved in the collaboration, the motivation for the collaboration, and measurable objectives resulting from this collaboration. Additionally, teams could collaborate to turn that need for collaboration into a ``x-as-a-service'' in order to avoid such need for collaboration in the future and, thus, such dependency. 

Once collaboration achieves significant results, structures such as platform teams become helpful. As innovations and improvements have been established, it is necessary to build platforms to reduce the cognitive load of product teams while maintaining their autonomy by consuming the platform. M. Pais argues that platform teams should ``interact" with product teams to define the platform itself. Consequently, the platform team is a product team whose customers are the internal product teams, which generates a great overload if we add other tasks, such as consulting, training, evangelization, and mentoring.

What M. Pais suggested is the importance of being clear about the type of interactions and behaviors expected, whatever the team may be and the interaction. This means, he believes there is a need to clearly define the types of interactions and expected behaviors, regardless of the team and the interaction. For example, a platform team can help product teams to be aligned to the flow, which makes sense in many cases because of  knowledge on the platform domain that the former owns. However, these platform teams must adopt the right behaviors and make an effort in these activities. Additionally, an enabler team could collaborate with a platform team to help define a new service based on the needs of product teams. There are ``typical'' types of interactions for each type of team, but it is not a rule, just the more common. The types of interactions aim to increase the adaptability of the organization. They should not be fixed or exclusively attributed to specific teams.

This feedback leads us to update T3 with interaction types and the following suggested attributes: duration, frequency, purpose, validation, dedication, and quality, but above all, to understand the importance of the interaction types and its dynamism to adapt to the teams maturity during their evolution adopting DevOps. 


\section{Threats to Validity, Reliability and Limitations}
\label{sec:threats}

Criteria for judging the quality of research designs are essential to establish the validity (i.e., the accuracy of the findings), and the reliability (i.e., the consistency of procedures and the researcher’s approach) of most empirical research \cite{creswell:2017,yin:2018}.

\subsection{Validity}

As described in Section~\ref{sec:methodology}, the research operation side of the theory research cycle is addressed from a qualitative point of view, recognizing multiple realities, as the constructivism philosophy states. This epistemological position makes assessing validity more complex (\cite{Shull:2008}, p. 306). Many researchers who adopt this stance believe that validity is too positivist and does not accurately reflect the nature of qualitative research. That is, as the constructivist stance assumes that reality is ``multiple and constructed”, then repeatability is simply not possible (\cite{Sandelowski:1993}). Attempts to develop frameworks to evaluate the contribution of constructivist research have encountered mixed reactions. We considered the quality criteria defined by Lincoln and Guba’s \cite{lincoln:1985} for qualitative research:

\textit{1. Credibility} is also referred to as trustworthiness, i.e., the extent to which conclusions are supported by rich, multivocal evidence. Lincoln and Guba’s \cite{lincoln:1985} presented some activities to mitigate this threat, such as those activities that make research more likely that credible findings and interpretations will be produced (prolonged engagement, persistent observation, and triangulation).


In this research, we focused on triangulation. The basic idea is to gather different types of evidence to support a proposition (\cite{Shull:2008}, p. 52). Denzin \cite{Denzin:1978} has suggested that four modes of triangulation exist: the use of multiple and different sources, methods, investigators, and theories. In this research, we collected data from multiple sources: i) the book by M. Skelton and M. Pais and the interview with Manuel Pais. 
We used different methods to collect these data (interview and literature) and different methods to analyze them: content analysis and grounded theory. Finally, we analyzed these data dually in two teams of researchers.

    \textit{2. Resonance} is the extent to which a study’s conclusions make sense to (resonate with) participants. A key strategy to that end is member checking so that some of the participants receive the preliminary results to ensure the correctness of our findings. Two teams of researchers participated in our study to collect, and process data from the different sources used and exchanged data between them to validate whether the interpretation of both teams on the same data set was coincident (as it was).

\textit{3. Usefulness} is the extent to which a study provides actionable recommendations to researchers, practitioners, or educators and the degree to which results extend our cumulative knowledge. This research has increased the scope of validity of our theory as far as the confirmation of the hypotheses is concerned.

\textit{4. Transferability} shows whether the findings could plausibly apply to other situations. Our theory has been operationalized and tested. That is, a set of hypotheses has been contrasted with the data of a case study (an interview and literature). The procedure used for testing can be replicated (transferred) when working with other (documentary) sources.

\textit{5. Conformability} assesses whether the findings emerge from the data collected from cases, not from preconceptions. The operationalization of the theory has been limited to its elements. Only the semantics of some indicators have been qualified to avoid more open interpretations. The data to confirm/refute the hypotheses were obtained from the case study described in Section~\ref{sec:usecase}.

\textit{6. Dependability} shows that the research process is systematic, well-documented, and can be traced. The public repository contains all the data and procedures used in this research so that other researchers can replicate it. This criterion is closely related to reliability.

\subsection{Reliability}

Reliability indicates that the researchers' approach is consistent across different researchers and projects \cite{creswell:2017}. During the case study's data analysis, we conducted collaborative qualitative analysis, specifically collaborative open coding. The benefits from collaborative coding have been extensively reported in \cite{Cornish:2014,hall:2005,Richards:2018} and include: (i) juxtaposing and integrating multiple and diverse perspectives (in our case, senior researchers with junior researchers and researchers from two different and geographically distant research groups with diverse backgrounds), which is often viewed as a way to counteract individual biases and enhance/increase credibility \cite{olson:2016}, trustworthiness \cite{patton:1999}, and rigor \cite{dube:2003}; (ii) addressing large and complex problems \cite{hall:2005} by effective management of large datasets \cite{olson:2016} (in our case, we had to manage more than 110 codes derived from the analysis of the collected data).; and (iii) effective mentoring of junior researchers \cite{Cornish:2014}. In Patton's words, ``Having two or more researchers independently analyze the same qualitative data set and then compare their findings provides an important check on selective perception, and blind interpretive bias" \cite{patton:1999}.

\subsection{Limitations}
The limitations of this research refer mostly to the scope of validity of the case study. Since statistical generalization is not plausible in qualitative research, the results of hypothesis testing should be limited to the case study analyzed. However, any refuted hypotheses should disappear from the corpus of the theory while those that have not been refuted remain. In this respect, we recall that ``A hypothesis cannot be proven, it can only be supported or refuted,...'' (\cite{Shull:2008}, 52).

\section{Conclusion}
\label{sec:conclusion}

In this work, we made a step forward in testing and validating theories within software engineering (SE). The contribution described in this paper is part of a broader research on building a theory. Previously~\cite{diaz2022}, we developed a unified Team Taxonomies Theory (T3) from 11 works that discuss and present various taxonomies of DevOps teams. In the present paper, we propose a methodological framework for theory-building with a continuous process of refinement and improvement. It consists of (conceptual) development, operationalization, confirmation or disconfirmation (testing), and application, being Sj{\o}berg et al. \cite{sjoberg:2008} the first authors in describing and adapting a theory-building framework to build theories in SE. However, most SE theorists only implement the first phase, that is, theory development, so that operationalization and testing, most of the time, are missing, or at least, they are not published, and there is no evidence for this in the current SE literature. In this paper, we present a confirmatory case study to perform the stages of operationalization and confirmation or disconfirmation (testing) of T3.

The goal of the theory operationalization phase was to derive the original 28 propositions of T3 into 34 testable hypotheses. To test these hypotheses, we conducted a confirmatory case study. As a result, we obtained 11 fully supported hypotheses, 3 partially tested hypotheses, and 20 non-tested hypotheses due to the lack of data and evidence. The most significant update in our theory  suggested by M. Pais is the need for separating platform building from those activities of facilitation, coaching, and support of other teams. This potential theory adjustment and the others mapped will be evaluated after more interview rounds. Therefore, it is necessary to continue the process of empirical induction, through the analysis of confirmatory case studies, which add more evidence in favor or against the hypotheses of the current theory.

At this time, the theory is ready to evolve and incorporate some aspects mentioned in the discussion that the initial theory did not address (e.g., adding more attributes and interaction modes in addition to collaboration). Finally, an effort in an explicable theory is necessary, that is, for the theory to be useful and usable by professionals of different profiles (from CEOs, managers to developers), it is necessary to move from formal models of the theory to more visual notations.   


In the near future, we will extend the scope of validity of the developed theory (through other confirmatory case studies) and, if necessary, broaden it to accommodate those aspects obtained during future empirical research that may not have been taken into account at present.

\section*{Data Availability}
\label{sec:dataavailability}
All the data related to this research, providing a detail chain of evidence, is available at \url{https://github.com/alvesisaque/devops_taxonomies.github.io}.


%
 \section*{Conflict of interest}
 The authors declare that they have no conflict of interest.

\bibliographystyle{spmpsci}      

\bibliography{bibliography}

\appendix
\section{Appendix Hypotheses}
\label{app:propositions}

Each variable is assigned a sequence number that will later be used as an identifying element within a precedence diagram. Empty cells indicate that there is no relationship between the variables in the corresponding row and column. If the cell is not empty, express the type of relation and, in brackets, the proposition number from which the relation is derived. For example, in the cell corresponding to the independent variable (column) $RO.S$ and the dependent variable (row) $Collaboration$ is labelled with $enable$ and $(1)$ indicating the type of relationship and coming from proposition 1 of the theory, which states: \textit{“A team culture based on responsibility/ownership sharing enables collaboration”}. 


\begin{sidewaystable}
\centering
\caption{Hypotheses generated. (1) Collaboration, (2)Responsibility ownership sharing, (3)Silos, (4)Automated Application Life\_Cycle Management, (5)Skills/Knowledge Sharing, (6)Cross-Functionality/skills Team, (7)Role Definition/Attributions, (8)Alignment of dev\&ops goals, (9)Metrics/visibility/feedback, (10)Enabler Team, (11)Bridge Team, (12)Self-Organization\&Autonomy Team, (13)Communications, (14)Transfer of Work, (15)Automated Infrastructure Management, (16)Platform Servicing}
\begin{tabular}{l|llllllllllllllll}
\cline{2-17}
 & \multicolumn{16}{c}{\textbf{Independent variables}} \\ \cline{2-17} 
 & \multicolumn{1}{l|}{1} & \multicolumn{1}{l|}{2} & \multicolumn{1}{l|}{3} & \multicolumn{1}{l|}{4} & \multicolumn{1}{l|}{5} & \multicolumn{1}{l|}{6} & \multicolumn{1}{l|}{7} & \multicolumn{1}{l|}{8} & \multicolumn{1}{l|}{9} & \multicolumn{1}{l|}{10} & \multicolumn{1}{l|}{11} & \multicolumn{1}{l|}{12} & \multicolumn{1}{l|}{13} & \multicolumn{1}{l|}{14} & \multicolumn{1}{l|}{15} & \multicolumn{1}{l|}{16} \\ \hline
\multicolumn{1}{|l|}{1} & \multicolumn{1}{l|}{} & \multicolumn{1}{l|}{enables(1)} & \multicolumn{1}{l|}{\cellcolor[HTML]{FFFFFF}} & \multicolumn{1}{l|}{impacts(3)} & \multicolumn{1}{l|}{enables(4)} & \multicolumn{1}{l|}{increase(5)} & \multicolumn{1}{l|}{decrease(6)} & \multicolumn{1}{l|}{\cellcolor[HTML]{FFFFFF}} & \multicolumn{1}{l|}{enables(8)} & \multicolumn{1}{l|}{} & \multicolumn{1}{l|}{} & \multicolumn{1}{l|}{} & \multicolumn{1}{l|}{} & \multicolumn{1}{l|}{} & \multicolumn{1}{l|}{} & \multicolumn{1}{l|}{} \\ \hline
\multicolumn{1}{|l|}{2} & \multicolumn{1}{l|}{\cellcolor[HTML]{FFFFFF}} & \multicolumn{1}{l|}{} & \multicolumn{1}{l|}{} & \multicolumn{1}{l|}{enables(16)} & \multicolumn{1}{l|}{} & \multicolumn{1}{l|}{} & \multicolumn{1}{l|}{} & \multicolumn{1}{l|}{} & \multicolumn{1}{l|}{} & \multicolumn{1}{l|}{enables(11)} & \multicolumn{1}{l|}{} & \multicolumn{1}{l|}{} & \multicolumn{1}{l|}{} & \multicolumn{1}{l|}{} & \multicolumn{1}{l|}{enables(15)} & \multicolumn{1}{l|}{} \\ \hline
\multicolumn{1}{|l|}{3} & \multicolumn{1}{l|}{reduces(2)} & \multicolumn{1}{l|}{reduces(10)} & \multicolumn{1}{l|}{} & \multicolumn{1}{l|}{} & \multicolumn{1}{l|}{} & \multicolumn{1}{l|}{reduces(19)} & \multicolumn{1}{l|}{} & \multicolumn{1}{l|}{} & \multicolumn{1}{l|}{} & \multicolumn{1}{l|}{} & \multicolumn{1}{l|}{become(11)} & \multicolumn{1}{l|}{} & \multicolumn{1}{l|}{} & \multicolumn{1}{l|}{} & \multicolumn{1}{l|}{} & \multicolumn{1}{l|}{} \\ \hline
\multicolumn{1}{|l|}{4} & \multicolumn{1}{l|}{impacts(3)} & \multicolumn{1}{l|}{} & \multicolumn{1}{l|}{difficult(21)} & \multicolumn{1}{l|}{} & \multicolumn{1}{l|}{} & \multicolumn{1}{l|}{increase(20)} & \multicolumn{1}{l|}{} & \multicolumn{1}{l|}{} & \multicolumn{1}{l|}{enables(22)} & \multicolumn{1}{l|}{provides(28)} & \multicolumn{1}{l|}{} & \multicolumn{1}{l|}{} & \multicolumn{1}{l|}{} & \multicolumn{1}{l|}{} & \multicolumn{1}{l|}{} & \multicolumn{1}{l|}{} \\ \hline
\multicolumn{1}{|l|}{5} & \multicolumn{1}{l|}{\cellcolor[HTML]{FFFFFF}} & \multicolumn{1}{l|}{} & \multicolumn{1}{l|}{} & \multicolumn{1}{l|}{enables(23)} & \multicolumn{1}{l|}{} & \multicolumn{1}{l|}{} & \multicolumn{1}{l|}{} & \multicolumn{1}{l|}{} & \multicolumn{1}{l|}{} & \multicolumn{1}{l|}{} & \multicolumn{1}{l|}{} & \multicolumn{1}{l|}{} & \multicolumn{1}{l|}{} & \multicolumn{1}{l|}{} & \multicolumn{1}{l|}{} & \multicolumn{1}{l|}{} \\ \hline
\multicolumn{1}{|l|}{6} & \multicolumn{1}{l|}{\cellcolor[HTML]{FFFFFF}} & \multicolumn{1}{l|}{\begin{tabular}[c]{@{}l@{}}is a property\\  of(9)\end{tabular}} & \multicolumn{1}{l|}{} & \multicolumn{1}{l|}{} & \multicolumn{1}{l|}{\begin{tabular}[c]{@{}l@{}}Is a property\\  of(17)\end{tabular}} & \multicolumn{1}{l|}{} & \multicolumn{1}{l|}{} & \multicolumn{1}{l|}{} & \multicolumn{1}{l|}{} & \multicolumn{1}{l|}{} & \multicolumn{1}{l|}{} & \multicolumn{1}{l|}{} & \multicolumn{1}{l|}{} & \multicolumn{1}{l|}{} & \multicolumn{1}{l|}{} & \multicolumn{1}{l|}{} \\ \hline
\multicolumn{1}{|l|}{7} & \multicolumn{1}{l|}{\cellcolor[HTML]{FFFFFF}} & \multicolumn{1}{l|}{} & \multicolumn{1}{l|}{} & \multicolumn{1}{l|}{} & \multicolumn{1}{l|}{} & \multicolumn{1}{l|}{} & \multicolumn{1}{l|}{} & \multicolumn{1}{l|}{} & \multicolumn{1}{l|}{} & \multicolumn{1}{l|}{} & \multicolumn{1}{l|}{} & \multicolumn{1}{l|}{} & \multicolumn{1}{l|}{} & \multicolumn{1}{l|}{} & \multicolumn{1}{l|}{} & \multicolumn{1}{l|}{} \\ \hline
\multicolumn{1}{|l|}{8} & \multicolumn{1}{l|}{requires(7)} & \multicolumn{1}{l|}{} & \multicolumn{1}{l|}{} & \multicolumn{1}{l|}{} & \multicolumn{1}{l|}{} & \multicolumn{1}{l|}{} & \multicolumn{1}{l|}{} & \multicolumn{1}{l|}{} & \multicolumn{1}{l|}{} & \multicolumn{1}{l|}{} & \multicolumn{1}{l|}{} & \multicolumn{1}{l|}{} & \multicolumn{1}{l|}{} & \multicolumn{1}{l|}{} & \multicolumn{1}{l|}{} & \multicolumn{1}{l|}{} \\ \hline
\multicolumn{1}{|l|}{9} & \multicolumn{1}{l|}{} & \multicolumn{1}{l|}{} & \multicolumn{1}{l|}{} & \multicolumn{1}{l|}{} & \multicolumn{1}{l|}{} & \multicolumn{1}{l|}{} & \multicolumn{1}{l|}{} & \multicolumn{1}{l|}{} & \multicolumn{1}{l|}{} & \multicolumn{1}{l|}{} & \multicolumn{1}{l|}{} & \multicolumn{1}{l|}{} & \multicolumn{1}{l|}{} & \multicolumn{1}{l|}{} & \multicolumn{1}{l|}{} & \multicolumn{1}{l|}{} \\ \hline
\multicolumn{1}{|l|}{10} & \multicolumn{1}{l|}{} & \multicolumn{1}{l|}{} & \multicolumn{1}{l|}{} & \multicolumn{1}{l|}{} & \multicolumn{1}{l|}{} & \multicolumn{1}{l|}{\begin{tabular}[c]{@{}l@{}}Is a property\\  of(18)\end{tabular}} & \multicolumn{1}{l|}{} & \multicolumn{1}{l|}{} & \multicolumn{1}{l|}{} & \multicolumn{1}{l|}{} & \multicolumn{1}{l|}{} & \multicolumn{1}{l|}{} & \multicolumn{1}{l|}{} & \multicolumn{1}{l|}{} & \multicolumn{1}{l|}{} & \multicolumn{1}{l|}{} \\ \hline
\multicolumn{1}{|l|}{11} & \multicolumn{1}{l|}{} & \multicolumn{1}{l|}{} & \multicolumn{1}{l|}{} & \multicolumn{1}{l|}{} & \multicolumn{1}{l|}{} & \multicolumn{1}{l|}{} & \multicolumn{1}{l|}{} & \multicolumn{1}{l|}{} & \multicolumn{1}{l|}{} & \multicolumn{1}{l|}{} & \multicolumn{1}{l|}{} & \multicolumn{1}{l|}{} & \multicolumn{1}{l|}{} & \multicolumn{1}{l|}{} & \multicolumn{1}{l|}{} & \multicolumn{1}{l|}{} \\ \hline
\multicolumn{1}{|l|}{12} & \multicolumn{1}{l|}{} & \multicolumn{1}{l|}{\begin{tabular}[c]{@{}l@{}}Is a property\\  of(12)\end{tabular}} & \multicolumn{1}{l|}{} & \multicolumn{1}{l|}{} & \multicolumn{1}{l|}{} & \multicolumn{1}{l|}{} & \multicolumn{1}{l|}{} & \multicolumn{1}{l|}{} & \multicolumn{1}{l|}{} & \multicolumn{1}{l|}{enables(24)} & \multicolumn{1}{l|}{} & \multicolumn{1}{l|}{} & \multicolumn{1}{l|}{} & \multicolumn{1}{l|}{} & \multicolumn{1}{l|}{} & \multicolumn{1}{l|}{} \\ \hline
\multicolumn{1}{|l|}{13} & \multicolumn{1}{l|}{} & \multicolumn{1}{l|}{enables(13)} & \multicolumn{1}{l|}{} & \multicolumn{1}{l|}{} & \multicolumn{1}{l|}{} & \multicolumn{1}{l|}{} & \multicolumn{1}{l|}{} & \multicolumn{1}{l|}{} & \multicolumn{1}{l|}{} & \multicolumn{1}{l|}{} & \multicolumn{1}{l|}{} & \multicolumn{1}{l|}{} & \multicolumn{1}{l|}{} & \multicolumn{1}{l|}{} & \multicolumn{1}{l|}{} & \multicolumn{1}{l|}{} \\ \hline
\multicolumn{1}{|l|}{14} & \multicolumn{1}{l|}{} & \multicolumn{1}{l|}{disables(14)} & \multicolumn{1}{l|}{} & \multicolumn{1}{l|}{} & \multicolumn{1}{l|}{} & \multicolumn{1}{l|}{} & \multicolumn{1}{l|}{} & \multicolumn{1}{l|}{} & \multicolumn{1}{l|}{} & \multicolumn{1}{l|}{} & \multicolumn{1}{l|}{} & \multicolumn{1}{l|}{} & \multicolumn{1}{l|}{} & \multicolumn{1}{l|}{} & \multicolumn{1}{l|}{} & \multicolumn{1}{l|}{} \\ \hline
\multicolumn{1}{|l|}{15} & \multicolumn{1}{l|}{} & \multicolumn{1}{l|}{} & \multicolumn{1}{l|}{} & \multicolumn{1}{l|}{} & \multicolumn{1}{l|}{} & \multicolumn{1}{l|}{} & \multicolumn{1}{l|}{} & \multicolumn{1}{l|}{} & \multicolumn{1}{l|}{} & \multicolumn{1}{l|}{} & \multicolumn{1}{l|}{} & \multicolumn{1}{l|}{} & \multicolumn{1}{l|}{} & \multicolumn{1}{l|}{} & \multicolumn{1}{l|}{} & \multicolumn{1}{l|}{} \\ \hline
\multicolumn{1}{|l|}{16} & \multicolumn{1}{l|}{} & \multicolumn{1}{l|}{} & \multicolumn{1}{l|}{} & \multicolumn{1}{l|}{is a(26)} & \multicolumn{1}{l|}{} & \multicolumn{1}{l|}{} & \multicolumn{1}{l|}{} & \multicolumn{1}{l|}{} & \multicolumn{1}{l|}{} & \multicolumn{1}{l|}{provides(25)} & \multicolumn{1}{l|}{} & \multicolumn{1}{l|}{} & \multicolumn{1}{l|}{} & \multicolumn{1}{l|}{} & \multicolumn{1}{l|}{is a(27)} & \multicolumn{1}{l|}{} \\ \hline
\end{tabular}
\end{sidewaystable}

\begin{table}[]
\caption{Hypotheses generated.}
\label{tab:indicatorsAnex}
\begin{tabular}{p{12cm}}
\rowcolor[HTML]{B7C8F7} 
\textbf{13 hypotheses about collaboration} \\
\rowcolor[HTML]{E7E6E6} 
P1. A team culture based on \textbf{\textbf{responsibility/ownership sharing}} {[}2{]} \textit{enables} \textbf{collaboration} {[}1{]} \\
\begin{tabular}[c]{@{}p{12cm}@{}}H1.1 A team culture based on \textbf{responsibility/ownership sharing} \textit{enables} \textbf{daily collaboration} between team members \\ H1.2 A team culture based on \textbf{responsibility/ownership sharing} \textit{enables} \textbf{high quality collaboration} between team members\end{tabular} \\
\rowcolor[HTML]{E7E6E6} P2. \textbf{Collaboration} {[}1{]} promoting \textit{reduces} \textbf{organizational silos/conflicts} {[}3{]} \\
\begin{tabular}[c]{@{}p{12cm}@{}}H2.1 \textbf{High quality, day-to-day collaboration} \textit{reduces} \textbf{organizational silos} \\ H2.2 \textbf{High quality, day-to-day collaboration} \textit{reduces} \textbf{cultural silos}\end{tabular} \\
\rowcolor[HTML]{E7E6E6} 
\begin{tabular}[c]{@{}p{12cm}@{}}P3. \textbf{Automated application life-cycle management} {[}4{]} \textit{is associated with} \textbf{collaboration} {[}1{]}. \textbf{Collaboration} impacts \textbf{Automated application life-cycle management} and vice versa. Automation and \textbf{collaboration} mutually facilitate the adoption of the other, so they are complementary\end{tabular} \\
\begin{tabular}[c]{@{}p{12cm}@{}}H3.1 \textbf{Automated application life-cycle management} \textit{promotes} the \textbf{collaboration} \\ H3.2 \textbf{collaboration} \textit{promotes} the \textbf{Automated application life-cycle management}\end{tabular} \\
\rowcolor[HTML]{E7E6E6} 
P4. A team culture based on \textbf{knowledge sharing} {[}5{]} \textit{enables} \textbf{collaboration} {[}1{]} \\
\begin{tabular}[c]{@{}p{12cm}@{}}H4.1 High levels of \textbf{knowledge sharing} (e.g., developers may have knowledge about\\ infrastructure/platform) \textit{enables} \textbf{high-quality, day-to-day collaboration} between teams \\  H4.2 Medium levels of \textbf{knowledge sharing} or minimal awareness of what is happening \\ on the other side of the wall, only allows for eventual and low-quality \textbf{collaboration} between teams\end{tabular} \\
\rowcolor[HTML]{E7E6E6} 
P5. If a team is characterized by \textbf{cross-functionality/skills} {[}6{]} this will \textit{increase} \textbf{collaboration} {[}1{]} \\
\begin{tabular}[c]{@{}p{12cm}@{}}H5.1 \textbf{Multidisciplinary/poly-skilled} teams (i.e., teams with all the necessary skills such as development, infrastructure, etc.) \textit{support} a better-quality \textbf{collaboration} with other teams \\ than those other teams with a lack of skills/knowledge/background \\ H5.2 \textbf{Multidisciplinary/poly-skilled} teams (i.e., teams with all the necessary skills such as development, infrastructure, etc.) allow for more frequent \textbf{collaboration} with other teams \\ than other teams with a lack of skills/knowledge/background\end{tabular} \\
\rowcolor[HTML]{E7E6E6} 
\begin{tabular}[c]{@{}p{12cm}@{}}P6. \textbf{collaboration} {[}1{]} \textit{is a property of} teams in which \textbf{skills take precedence over roles}, \\ i.e., the role definition/attributions {[}7{]} code; hence, if there are already separate roles, \\ responsibilities are very clear and \textbf{collaboration} is not fostered or promoted\end{tabular} \\
H6.1 Teams with \textbf{well-defined and differentiated roles} may \textit{decrease} \textbf{collaboration} \\
\rowcolor[HTML]{E7E6E6} 
P7. A \textbf{collaboration-based} {[}1{]} culture \textit{requires} \textbf{alignment of dev \& ops goals} {[}8{]} \\
H7.1 \textbf{collaboration} \textit{requires} that dev\&ops teams to \textbf{share business goals} and global ones (product thinking) \\
\rowcolor[HTML]{E7E6E6} 
P8. A team culture based on \textbf{metrics/visibility/feedback} {[}9{]} \textit{enables} \textbf{collaboration} {[}1{]} \\
H8.1 Regularly performing the measurement and sharing activities reinforces the collaborative culture
\end{tabular}
\end{table}

\begin{table}[]
\begin{tabular}{p{12cm}}
\rowcolor[HTML]{B7C8F7} 
\textbf{9 hypotheses about \textbf{responsibility/ownership sharing}} \\
\rowcolor[HTML]{E7E6E6} 
P9. \textbf{responsibility/ownership sharing} {[}2{]} \textit{is a property of} \textbf{cross-functionality/skills} {[}6{]} teams \\
\begin{tabular}[c]{@{}p{12cm}@{}}H9.1 \textbf{cross-functionality/skills} teams are \textit{characterized by} sharing all or some responsibility of products, output artefacts (e.g., databases), and tasks (e.g., NFR shared responsibility, infrastructure management shared responsibility, monitoring shared responsibility, and incident handling shared responsibility, etc.) So that everyone is responsible for build, test, deploy, operate and maintain their applications and infrastructure (although some responsibilities may still reside on one side -dev or ops-)\end{tabular} \\
\rowcolor[HTML]{E7E6E6} 
P10. \textbf{responsibility/ownership sharing} {[}2{]} \textit{reduces} \textbf{organizational silos/conflicts} {[}3{]} \\
\begin{tabular}[c]{@{}p{12cm}@{}}H10.1 The greater the degree of sharing, the fewer the number of cultural silos (which have to do with practices and culture and less to do with team structure as is the case with organizational silos). This hypothesis can be derived by combining by modus ponendo ponens the hypotheses H1 \& H2\end{tabular} \\
\rowcolor[HTML]{E7E6E6} 
\begin{tabular}[c]{@{}p{12cm}@{}}P11. \textbf{responsibility/ownership sharing} {[}2{]} \textit{is a property of} organizational structures that rely on an \textbf{enabler (platform) team} {[}10{]}. The existence of platform teams does not lead to a separation of responsibilities but rather they become facilitators and make ownership sharing possible, unlike dev ops (bridge) teams that become new silos with their own responsibilities (e.g., deployment, monitoring, etc.).\end{tabular} \\
\begin{tabular}[c]{@{}p{12cm}@{}}H11.1 The existence of enabler (platform) teams become facilitators and make ownership sharing possible \\ H11.2 dev ops (bridge) teams {[}11{]} become new silos with their own responsibilities (e.g., deployment, monitoring, etc.)\end{tabular} \\
\rowcolor[HTML]{E7E6E6} 
P12. \textbf{responsibility/ownership sharing} {[}2{]} \textit{is a property of} team \textbf{self-organization \& autonomy} {[}12{]} \\
\begin{tabular}[c]{@{}p{12cm}@{}}H12.1 If a team in the \textbf{autonomy} attribute has the value self-organisation, this implies that there is full/medium \textbf{responsibility/ownership sharing}.\end{tabular} \\
\rowcolor[HTML]{E7E6E6} 
P13. A team culture based on \textbf{responsibility/ownership sharing} {[}2{]} \textit{enables} \textbf{communication} {[}13{]} \\
\begin{tabular}[c]{@{}p{12cm}@{}}H13.1 A team culture base on full/medium \textbf{responsibility/ownership sharing} (a shared vocabulary also emerged from sharing and this facilitates communication) \textit{enables} a \textbf{frequent communication}\end{tabular} \\
\rowcolor[HTML]{E7E6E6} 
\begin{tabular}[c]{@{}p{12cm}@{}}P14. \textbf{responsibility/ownership sharing} {[}2{]} \textit{is associated with} the \textbf{transfer of work} between teams {[}14{]}. If there is no shared responsibility, there is necessarily a transfer of work between development to production and operation teams (and vice versa)\end{tabular} \\
\begin{tabular}[c]{@{}p{12cm}@{}}H14.1 If dev and ops have separate responsibilities and tasks (each team member has different responsibilities and tasks, i.e. \textbf{responsibility/ownership sharing}: null\_sharing) there is necessarily a transfer of work between development to production and operation teams (and vice versa)\end{tabular} \\
\rowcolor[HTML]{E7E6E6} 
P15. \textbf{Automated infrastructure management} {[}15{]} \textit{enables} \textbf{responsibility/ownership sharing} {[}2{]}. \\
\begin{tabular}[c]{@{}p{12cm}@{}}H15.1 If Ops develops IaC, it performs Dev tasks which enhances \textbf{responsibility/ownership sharing} (from null\_sharing to full/medium sharing)\end{tabular} \\
\rowcolor[HTML]{E7E6E6} 
P16. \textbf{Automated application life-cycle management} {[}4{]} \textit{enables} \textbf{responsibility/ownership sharing} {[}2{]} \\
H16.1 \textbf{Automated application life-cycle management} encourages the move from null\_sharing to full/medium sharing
\end{tabular}
\end{table}

\begin{table}[]
\begin{tabular}{p{12cm}}
\rowcolor[HTML]{B7C8F7} 
\textbf{4 hypotheses about \textbf{cross-functionality/skills}} \\
\rowcolor[HTML]{E7E6E6} 
P17. \textbf{Skills/knowledge sharing} {[}5{]} \textit{is a property of} teams characterized by \textbf{cross-functionality/skills} {[}6{]}. \\
\begin{tabular}[c]{@{}p{12cm}@{}}H17.1 The \textbf{cross-functionality/skills} teams also are \textit{characterised by} high o medium level of \textbf{skills/knowledge sharing} (e.g., developers may have knowledge about infrastructure/platform)\end{tabular} \\
\rowcolor[HTML]{E7E6E6} 
P18. \textbf{cross-functionality/skills} {[}6{]} \textit{is a property of} enabler (platform) team {[}10{]}. \\
H18.1 \textbf{Enabler teams} need a variety of skills to develop a platform. \\
\rowcolor[HTML]{E7E6E6} 
P19. \textbf{cross-functionality/skills} {[}6{]} reduces \textbf{organizational silos/conflicts} {[}3{]} \\
\begin{tabular}[c]{@{}p{12cm}@{}}H19.1 \textbf{Multidisciplinary/poly-skilled} teams (i.e., teams with all the necessary skills such as development, infrastructure, etc.) \textit{avoid} \textbf{organizational silos}\end{tabular} \\
\rowcolor[HTML]{E7E6E6} 
P20. If a team is characterized by \textbf{cross-functionality/skills} {[}6{]} this will \textit{increase} \textbf{automated application life-cycle}
\textbf{management} {[}4{]} \\
\rowcolor[HTML]{FFFFFF} 
H20.1 \textbf{Multidisciplinary/poly-skilled} teams will \textit{increase} \textbf{Automated application life-cycle management}
\end{tabular}
\end{table}

\begin{table}[]
\begin{tabular}{p{12cm}}
\rowcolor[HTML]{B7C8F7} 
\textbf{3 hypotheses about \textbf{Automated application life-cycle management}} \\
\rowcolor[HTML]{E7E6E6} 
P21. \textbf{organizational silos/conflicts} {[}3{]} make the adoption of an \textbf{Automated application life-cycle management} {[}4{]} difficult \\
H21.1 \textbf{organizational silos/conflicts} make the adoption of an \textbf{Automated application life-cycle management} difficult \\
\rowcolor[HTML]{E7E6E6} 
P22. Metrics, visibility \& feedback {[}9{]} \textit{enables} \textbf{Automated application life-cycle management} {[}4{]} \\
\begin{tabular}[c]{@{}p{12cm}@{}}H22.1 Metrics (process, value, cost, and technical metrics) provide fast and continuous feedback from users, reduce the risk and cost of deployments, get better visibility into the delivery process itself, and manage the risks of software delivery more effectively\end{tabular} \\
\rowcolor[HTML]{E7E6E6} 
P23. \textbf{Automated application life-cycle management} {[}4{]} \textit{enables} \textbf{skills/knowledge sharing} {[}5{]} \\
H23.1 \textbf{Automation} \textbf{increases} trust between teams and team sharing
\end{tabular}
\end{table}

\begin{table}[]
\begin{tabular}{p{12cm}}
\rowcolor[HTML]{B7C8F7} 
\textbf{3 hypothesis about enabler platform} \\
\rowcolor[HTML]{E7E6E6} 
P24. \textbf{Enabler (platform)} team {[}10{]} \textit{enables} team \textbf{self-organization \& autonomy} {[}12{]} \\
\begin{tabular}[c]{@{}p{12cm}@{}}H24.1 An \textbf{enabling team} \textit{is characterised} by \textbf{cross-functionality/skills} (H18.1) and this allows for autonomy: self-organisation (can access all the information needed to develop, deploy and operate the product) without having to wait for anyone else.\end{tabular} \\
\rowcolor[HTML]{E7E6E6} 
P25. \textbf{Enabler (platform)} team {[}10{]} \textit{provides} \textbf{platform servicing} {[}16{]} \\
H25.1 \textbf{Enabler team} \textit{provides} \textbf{platform servicing} to product teams to assist them on DevOps platform \\
\rowcolor[HTML]{E7E6E6} 
P28. \textbf{Enabler (platform)} teams {[}10{]} \textit{provide} \textbf{Automated application life-cycle management} {[}4{]} \\
\rowcolor[HTML]{FFFFFF} 
\begin{tabular}[c]{@{}p{12cm}@{}}H28.1 \textbf{Enabler (platform)} teams \textit{provide} \textbf{Automated application life-cycle management} which is nothing more than a special case of platform servicing\end{tabular} \\
\end{tabular}
\end{table}

\begin{table}[]
\begin{tabular}{p{12cm}}
\rowcolor[HTML]{B7C8F7} 
\textbf{2 relational hypothesis “is a”} \\
\rowcolor[HTML]{E7E6E6} 
P26. \textbf{Automated application life-cycle management} {[}4{]} \textit{is a} \textbf{Platform servicing} {[}16{]} \\
H26.1 \textbf{Automated application life-cycle management} \textit{is a} special case of \textbf{platform servicing} {[}16{]} \\
\rowcolor[HTML]{E7E6E6} 
P27. \textbf{Automated infrastructure management} {[}15{]} \textit{is a} Platform servicing {[}16{]} \\
H27.1 \textbf{Automated infrastructure management} \textit{is a} special case of \textbf{platform servicing}
\end{tabular}
\end{table}

\begin{table}[]

\begin{tabular}{p{6,2cm} p{6cm}}
\end{tabular}
\end{table}

\end{document}